\documentclass[apl,a4paper,twocolumn]{revtex4}  % for review and submission
\pdfoutput=1

\usepackage{graphicx}
\usepackage{ucs} % for unicode characters

\usepackage{color} % for nice colors in the links
\definecolor{darkblue}{rgb}{0,0,0.5}
\definecolor{lila}{rgb}{0.3,0,0.3}
\definecolor{turq}{rgb}{0,0.1,0.4}

\usepackage{natbib} % that hyperlinks show up in the bibliography
\usepackage{url} % that hyperlinks may be hyphenated correctly

\usepackage[pdftex,
colorlinks=true,
 linkcolor=darkblue, % usual links
 filecolor=red,
 citecolor=turq, % for bibliographic
 urlcolor=lila, % for Emails etc
 pdftitle={Integrated Diamond Optics for Single Photon Detection},
 pdfauthor={P. Siyushev and F. Kaiser and V. Jacques and I. Gerhardt and S. Bischof and H. Fedder and J. Dodson and M. Markham and D. Twitchen and F. Jelezko and J. Wrachtrup},
 pdfsubject={Integrated Diamond Optics for Single Photon Detection},
 pdfkeywords={Solid Immersion Lens, SIL, NV Centers, diamond optics, quantum information processing},
 pdfpagelabels=true,
 breaklinks=true,
 plainpages=false,
 bookmarks=false, bookmarksnumbered=true
]{hyperref}

\usepackage{graphicx}  % needed for figures
\usepackage{dcolumn}   % needed for some tables
\usepackage{bm}        % for math
\usepackage{amssymb}   % for math
\usepackage[english]{babel}

\begin{document}

\title{Integrated Diamond Optics for Single Photon Detection}

\date{\today}
\author{P.~Siyushev$^{1}$, F.~Kaiser$^{1}$, V.~Jacques$^{1}$, I.~Gerhardt$^{1,2}$, S.~Bischof$^{1}$, H.~Fedder$^{1}$,  J.~Dodson$^{3}$, M.~Markham$^{3}$, D.~Twitchen$^{3}$, F.~Jelezko$^{1}$ and J.~Wrachtrup$^{1,2}$}

\affiliation
{$^{1}$3.Physics Institute and Research Center SCOPE, University of Stuttgart, D-70550 Stuttgart, Germany\\
$^{2}$Max Planck Institute for Solid State Research, Heisenbergstra\ss{}e 1, D-70569 Stuttgart, Germany\\
$^{3}$Element Six Ltd., King's Ride Park, Ascot, Berkshire SL5 8BP, UK}
\begin{abstract}
Optical detection of single defect centers in the solid state is a key element of novel quantum technologies. This includes the generation of single photons and quantum information processing. Unfortunately the brightness of such atomic emitters is limited. Therefore we experimentally demonstrate a novel and simple approach that uses off-the-shelf optical elements. The key component is a solid immersion lens made of diamond, the host material for single color centers. We improve the excitation and detection of single emitters by one order of magnitude, as predicted by theory.
\end{abstract}
\maketitle
Over the last decade, solid-state optically-active defects have attracted attention owing to applications in quantum information science. Among numerous studied color centers in diamond, the negatively-charged nitrogen-vacancy (NV) defect is particularly interesting. For the NV-center a variety of applications has been demonstrated, such as diamond-based single photon sources~\cite{Kurtsiefer_PRL2000,Beveratos_PRA2001}, spin based magnetic field sensors~\cite{Gopi_Nature2008,Maze_Nature2008} and diamond-based quantum information protocols~\cite{Weber2010}. Its spin state can be coherently manipulated with high fidelity~\cite{Fuchs2009} owing to long coherence time~\cite{Gupi_NatMat2009}. Moreover, the spin of NV defects can be efficiently read out~\cite{Neumann_Science2010} and coupled to photons through spin-dependent transitions~\cite{Togan2010}, a key ingredient towards long-distance quantum communications using photons as flying qubits.

For all above mentioned experiments the photon collection efficiency is crucial. For single NV defects this is reduced by the high refractive index of diamond ($n_{d}=2.4$), owing to total internal reflection at the diamond-air interface. Therefore several approaches to increase the collection efficiency, such as optical elements like metallic nanoantenna~\cite{Benson_NanoLett2009} or photonic wire microcavities~\cite{Babinec_NatNano2010}, have been suggested. Improving light extraction from diamond can also be achieved by using NV defects hosted in nanocrystals with a size smaller than the emission wavelength. In this case, refraction becomes irrelevant and the defect can be considered as a point source emitting in the surrounding medium. However, the spin properties of nanocrystals are often affected by surface defects leading to a short coherence time. To overcome these limitations we engineer a solid immersion lens directly into the host material diamond~\cite{Mansfield_APL1990,diamondSILobrien}.

To estimate theoretically the collection efficiency, we assume a single emitter in diamond and use geometrical optics and Fresnel's reflection laws. For a single emitting dipole the normalized photoluminescence (PL) intensity $I_{s}$ (resp. $I_{p}$) having $s$ (resp. $p$) polarization can be calculated as~\cite{Zwiller_JAP2002}

\begin{eqnarray}
	I_s &=& \frac{3}{8 \, \pi} \left[1-\sin^2(\theta)\cos^2(\phi)\right] \sin^2(\phi) \qquad \mbox{and}\\
	I_p &=& \frac{3}{8 \, \pi} \left[1-\sin^2(\theta)\cos^2(\phi)\right] \cos^2(\phi) \ ,
    \label{dipoles}
\end{eqnarray}
where $\theta$ is the emission angle and $\phi$ the azimuthal angle measured from the dipole axis.

\begin{figure}[t]
\centerline
{\includegraphics[width=8.5cm]{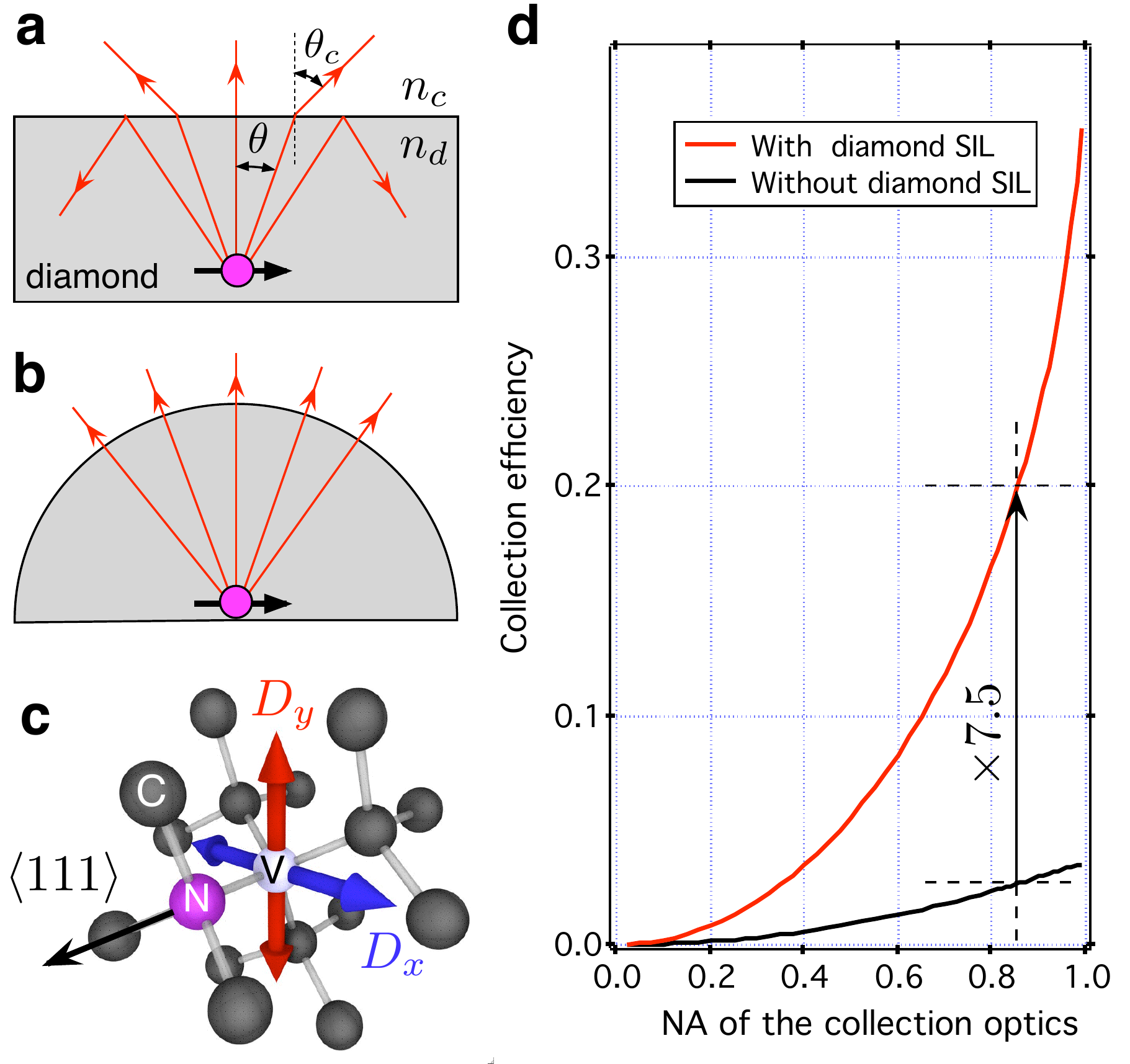}}
\caption{(a),(b) Ray propagation (a) for a single emitting dipole oriented parallel to the diamond flat surface and (b) for a dipole located at the center of an hemispherical diamond solid-immersion lens. (c) Atomic structure of the nitrogen-vacancy (NV) defect in diamond indicating two orthogonal dipoles $D_{x}$ and $D_{y}$, in the plane perpendicular to the NV symmetry axis ($\langle 111 \rangle$ crystalline axis). (d) Computed collection efficiencies of NV defect PL as a function of the numerical aperture (NA) of the collection optics. The simulation is performed for NV defects in a $[100]$-oriented sample with (red curve) and without the diamond SIL (blue curve). For a numerical aperture NA$=0.85$, a 7.5-fold enhancement of collection efficiency is expected while using a diamond SIL. For the simulation $n_{c}=1$ and $n_{d}=2.4$.}
\label{Fig1}
\end{figure}

We first consider a dipole beneath a planar diamond surface (Fig.~\ref{Fig1}(a)). For the moment we assume the dipole is oriented parallel to the surface. The more general case will be treated below. At the diamond interface, Snell's law imposes $n_{d}\sin\theta=n_{c}\sin\theta_{c}$, where $\theta_{c}$ is the propagation angle in the collection optics medium, characterized by a refractive index $n_{c}$. As a result, most of the PL is trapped inside the diamond matrix, since total internal reflection is achieved as soon as $\theta\geqslant \theta_{\rm TIR}=\arcsin(n_{c}/n_{d})$. For a diamond-air interface this limits the collection angle to $\theta_{\rm TIR}\approx 24.6^{\circ}$. The effective numerical aperture (NA) over which the dipole radiation will be integrated reads $\theta_{\rm m}=\arcsin({\rm NA}/n_{d})$.

The collection efficiencies $\eta_s$ and $\eta_p$ are obtained by integrating over the solid angle defined by the NA of the collection optics, leading to
\begin{equation}
	\eta_{s,p} = \int_{0}^{\theta_{\rm m}} \int_0^{2\pi} I_{s,p}  \, T_{s,p} \, \sin(\theta) \, d\phi\,d\theta
	\label{Eq1}
\end{equation}
\noindent where $T_s$ and $T_p$ are respectively the transmission coefficients of the $s$ and $p$ polarization components through the interface.

We now consider a dipole located at the center of a hemispherical diamond SIL used as primary collection optic (Fig.~\ref{Fig1}(b)). In that case, all light rays exit the SIL in a direction normal to the surface, thus avoiding any refraction at the interface. Note that thereby also all chromatic abberations are suppressed. As a result, the collection efficiencies $\eta_s$ and $\eta_p$ are defined by equation~(\ref{Eq1}), while using  $\theta_{\rm m}=\arcsin({\rm NA})$ and  $T_{s}=T_{p}=\frac{4n_{c}n_{d}}{(n_{c}+n_{d})^{2}}$.

In the general case the collection efficiency depends on the dipole orientation inside the diamond matrix. For NV defects, the PL is associated with two orthogonal dipoles $D_{x}$ and $D_{y}$ located in a plane perpendicular to the NV defect symmetry axis~\cite{Epstein_NatPhys2005}, which is a $\langle 111\rangle$ crystalline axis (Fig.~\ref{Fig1}(c)). In addition, the absolute orientation of the two transition dipoles is determined by the direction of non-axial local strain in the diamond matrix. Since local strain is randomly distributed over a diamond sample, the orientation of the two orthogonal dipoles is random in a plane perpendicular to the NV defect symmetry axis.

For a $[100]$-oriented diamond sample, the absolute angle between the NV defect axis and the diamond flat surface is the same for the four possible NV defect orientations~\footnote{The NV defect can be oriented along [$111$], [$\bar{1}\bar{1}1$], [$1\bar{1}1$], or [$\bar{1}11$] direction.}. To evaluate the collection efficiencies for this case, we take into account the orientation of the two dipoles with respect to the flat diamond surface and apply the required rotations to equation~(\ref{Eq1}). An average over the two dipoles and over all possible strain-induced orientations is taken. The results of the calculation are shown on Fig.~\ref{Fig1}(d). Using a $0.85$-NA microscope objective, the collection efficiency is $2.6\%$ for a planar diamond surface and reaches $19.6\%$ with a hemispherical diamond SIL. From this geometrical considerations, a 7.5-fold enhancement in collection efficiency is therefore expected for that particular NA of the collection optics. Note that this theoretical value requires a diamond SIL with a surface roughness of approximatively $\lambda/10$~\cite{Baba_JAP1990}.

A diamond SIL was fabricated as follows. Homoepitaxial chemical vapor deposition (CVD) diamond growth was performed using a selected high pressure high temperature grown substrate, which had been carefully processed prior to growth in order to minimize the dislocation density in the grown CVD diamond.  A two stage chemical process was subsequently used to produce a grown CVD layer with a thickness of about 2.3~mm made up with a first layer containing $\sim$0.2~ppm of single substitutional nitrogen. The final layer of $\sim$0.6~mm was engineered with a significant reduced nitrogen concentration ($<5$~ppb), to reduce the number of native NV defects. The SIL was subsequently processed from this CVD diamond sample, the flat surface being composed by the high purity layer described above, oriented along a $[100]$ crystalline axis.  Using a combination of laser and mechanical processing stages~\footnote{Nelissen, W. {\it et al.}, Patent WO 2007/007126A1, international publication date 18/01/2007.}, a 1-mm diameter SIL was produced with flatness better than 10~nm (rms), measured by atomic force microscopy.

\begin{figure}[t]
\centerline
{\includegraphics[width=8.5cm]{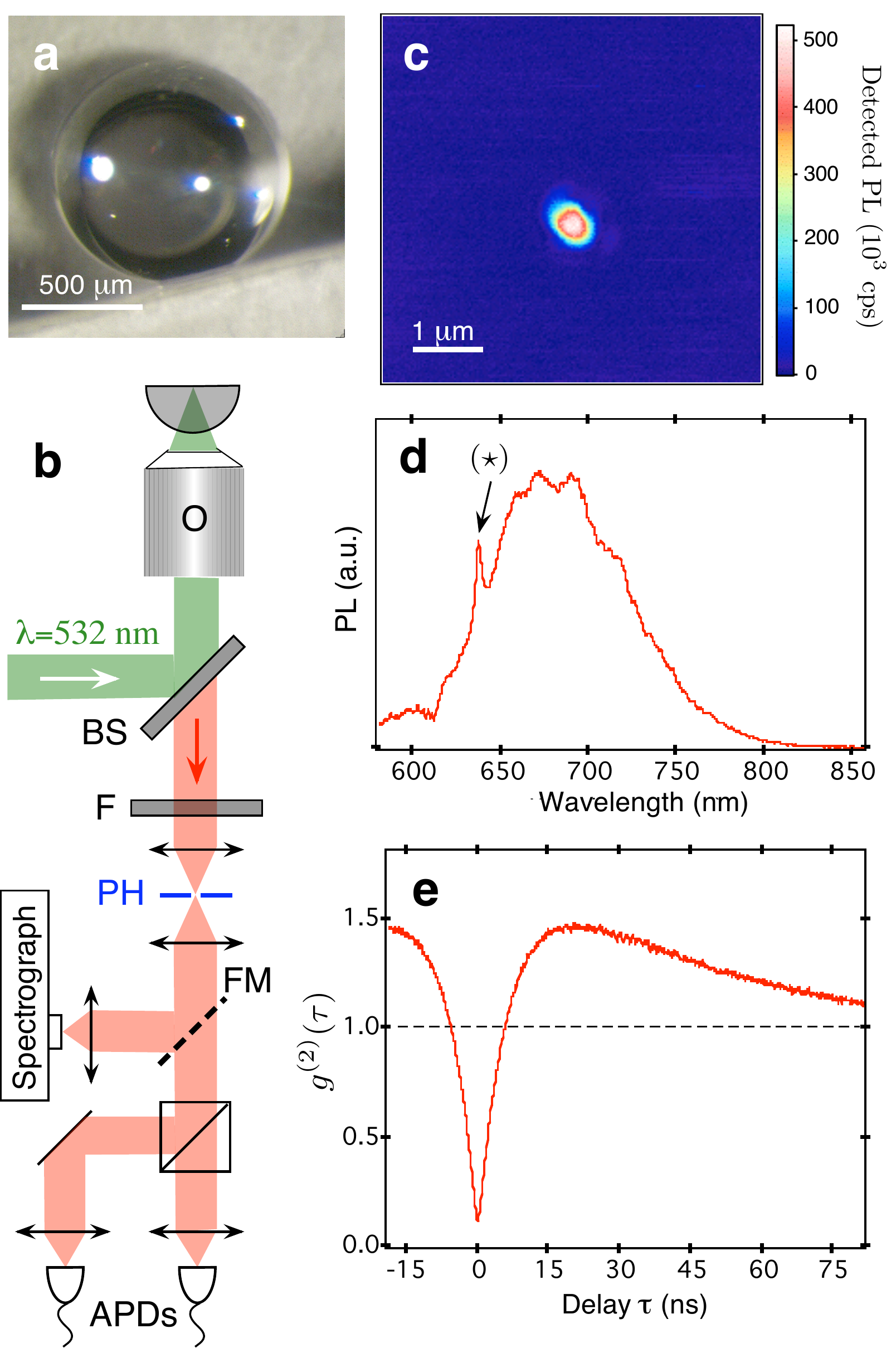}}
\caption{(a) Picture of the diamond SIL. (b) Experimental setup. BS: quartz plate with $5\%$ reflectance; O: microscope objective with a $0.85$ numerical aperture; PH: $50 \ \mu$m  diameter pinhole; F: $580$ nm long-pass filter ; FM: flip mirror directing the collected PL either to an imaging spectrometer (Acton research) or to a Hanbury-Brown and Twiss interferometer consisting of two silicon avalanche photodiodes (APD) placed on the output ports of a $50/50$ beamsplitter. (c) Typical raster scan of a region close to the center of the diamond SIL for a laser power of $200 \ \mu$W. The bright spot indicates a single NV defect, imaged with a signal to background ratio of $100$:$1$ . (d) PL spectrum showing a broad emission with a zero phonon line at the wavelength $\lambda=637$ nm. (e) Second-order autocorrelation function $g^{(2)}(\tau)$.}
\label{Fig2}
\end{figure}

Native single NV defects in the diamond SIL were imaged using conventional confocal microscopy, as depicted in figure~\ref{Fig2}(b). A frequency doubled Nd:YAG laser ($\lambda=532$~nm) was tightly focused onto the flat surface of the diamond SIL using a $0.85$-NA microscope objective. The PL was collected by the same objective using the SIL as primary collection optics, and spectrally filtered from the remaining excitation light. The collected light was then focused onto a $50 \ \mu$m diameter pinhole and directed either to a spectrometer or to a Hanbury-Brown and Twiss (HBT) interferometer for photon correlation measurements. A PL raster scan of a region close to the center of the diamond SIL is depicted in figure~\ref{Fig1}(b), showing an isolated photoluminescent spot. The spectrum recorded at this point exhibits a broadband emission with a characteristic zero-phonon line at the wavelength $\lambda_{\rm ZPL}=637$ nm (Fig~\ref{Fig2}(d)) which is associated with an NV defect. We note that optical aberrations can be clearly observed in the confocal raster scan, since the shape of the bright spot is not perfectly circular. Such aberrations are attributed to a lateral offset of $10 \ \mu$m between the NV defect location and the center of the diamond SIL~\cite{Baba_JAP1990}.

To proof that research was performed on a single emitter, we measured the time delay histogram between two consecutive single-photon detections using a standard HBT interferometer. After normalization to a Poissonnian statistics, the recorded histogram is equivalent to a measurement of the second-order autocorrelation function $g^{(2)}(\tau)$ defined by
\begin{equation}
g^{(2)}(\tau)=\frac{\langle I(t)I(t+\tau)\rangle}{\langle I(t)\rangle^{2}} \ ,
\label{defg2}
\end{equation}
where $I(t)$ is the PL intensity at time $t$. As shown in Fig.~\ref{Fig2}(e), a sub-poissonian photon statistics ($g^{(2)}(0)\approx 0.1$) is observed, proving a single NV defect.

\begin{figure}[b]
\centerline
{\includegraphics[width=8.5cm]{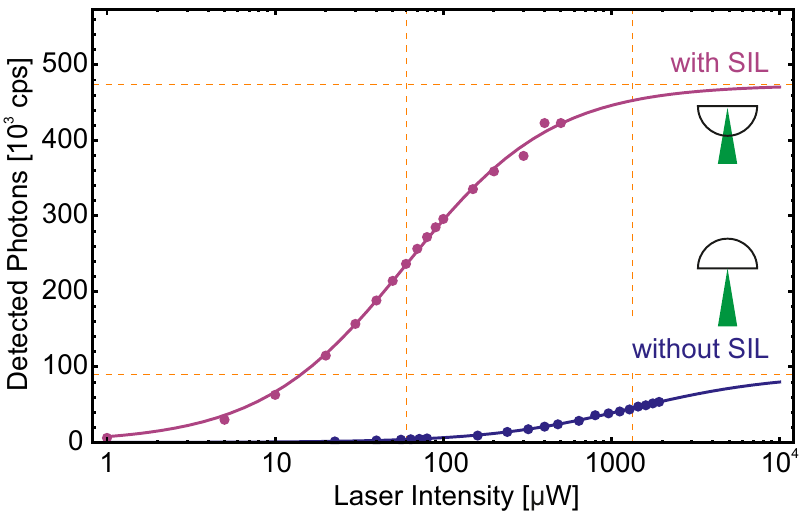}}
\caption{Background-corrected PL as a function of the laser intensity. Red and blue points correspond respectively to a single NV defect imaged through the diamond SIL and through a planar diamond surface. Solid lines are data fitting using equation~(\ref{sat}).}
\label{Fig3}
\end{figure}

With the aim of quantifying the enhancement in collection efficiency, the PL rate $R$ was measured as a function of the laser intensity $\mathcal{I}$ (Fig.~\ref{Fig3}). Experimental data were then fitted using a two-level model with the relation
\begin{equation}
R=R_{\infty}\frac{\mathcal{I}}{\mathcal{I}+\mathcal{I}_{\rm sat}} \ ,
\label{sat}
\end{equation}
where $\mathcal{I}_{\rm sat}$ is the saturation laser intensity for the given focus size and $R_{\infty}$ the emission rate at saturation. For the studied single NV defect imaged through the diamond SIL, we find $R_{\infty}^{\rm SIL}=493 \pm 5 \times10^3$~cps and $\mathcal{I}_{\rm sat}^{\rm SIL}=61\pm 1$~$\mu$W. At this saturation intensity the photon emission is enhanced by two orders of magnitude with respect to a collection in a bulk diamond. This we check by a single NV defect imaged from the backside of the SIL, {\it i.e.}~imaged through a planar diamond surface. In that configuration, the saturation parameters are $R_{\infty}^{\rm no SIL}=80\pm 2 \times10^3$~cps and $\mathcal{I}_{\rm sat}^{\rm no SIL}=1330\pm 30 \ \mu$W. A more than 6-fold enhancement is therefore achieved in high saturation conditions. Such a value is slightly smaller than the one predicted by pure geometrical optic arguments (Fig.~\ref{Fig1}(d)). This discrepancy is attributed to optical aberrations arising from unperfect positioning of the NV defect in the center of the diamond SIL~\cite{Baba_JAP1990}. We note that similar collection efficiency enhancements were obtained for two others single NV defects located around $10\ \mu$m away from the center of the SIL. However, if this distance exceeds $30 \ \mu$m, the enhancement vanishes. A precise positioning of NV defects at the center of the SIL can be achieved by controlled ion implantation~\cite{Meijer_APL2005}. In addition, collection efficiency could be further improved by using an anti-reflection coating on the hemispheric surface and/or a properly designed Bragg mirror deposited on the SIL flat surface, in order to collect backwards emitted photons.

Beyond collection efficiency enhancement, we note that exciting through the diamond SIL yields to a smaller excitation volume. Indeed, the saturation laser intensity $\mathcal{I}_{\rm sat}$ is decreased by roughly one order of magnitude while exciting a single NV defect through the diamond SIL. This effect might have interesting applications in the context of stimulated emission depletion (STED) microscopy, since the resolving power of STED is proportional to $\sqrt{\mathcal{I}_{\rm sat}/\mathcal{I}}$~\cite{Rittweger_NatPhot2009}. Furthermore, the reduced focus of the excitation laser funnels more efficiently photons towards the single emitter. This allows for the detection of a single defect in transmission through an efficient extinction of the excitation laser beam~\cite{Sandoghdar_NatPhys2008}.

\noindent \subsection*{Acknowledgements}
This work was supported by the EU, DFG, Landesstiftung BW, BMBF, Volkswagen Stiftung and NIH.

\noindent \subsection*{Additional Information}Patent application related to data present in this manuscript is under submission process. Correspondence and requests for materials should be addressed to V.J. (vincent.jacques@lpqm.ens-cachan.fr) and F.J. (f.jelezko@physik.uni-stuttgart.de).

\end{document}